\documentclass[aps,pra,twocolumn,superscriptaddress,unsortedaddress,nofootinbib,nobalancelastpage,10pt]{revtex4-2}

\usepackage{physics}
\usepackage{amssymb}
\usepackage{graphicx}
\usepackage{mathtools}
\usepackage[x11names]{xcolor}
\usepackage{xspace}
\usepackage{xparse}
\usepackage{hyperref}
\usepackage[per-mode=symbol]{siunitx}
\usepackage[normalem]{ulem}
\usepackage{cancel}

\hypersetup{
    colorlinks=true, 
    linkcolor=blue,  
    citecolor=magenta,  
    urlcolor=magenta    
}

\renewcommand{\t}[1]{t_\mathrm{#1}}
\newcommand{\spt}[1]{^\mathrm{#1}}
\newcommand{\sbt}[1]{_\mathrm{#1}}

\newcommand{\tgt}{t\sbt{gate}}
\newcommand{\trp}{t\sbt{ramp}}
\newcommand{\thd}{t\sbt{hold}}
\newcommand{\spL}{\spt{L}}
\newcommand{\spR}{\spt{R}}
\newcommand{\sbL}{\sbt{L}}
\newcommand{\sbR}{\sbt{R}}
\newcommand{\esp}{\epsilon\sbt{swap}}
\newcommand{\elk}{\epsilon\sbt{leak}}

\newcommand{\sqiswap}{$\sqrt{i\mathrm{SWAP}}$\xspace}
\newcommand{\vbeta}{$\vb*{V}^\mathrm{\beta}$\xspace}
\newcommand{\vzeta}{$\vb*{V}^\mathrm{\zeta}$\xspace}

\begin{document}

\title{Design and Dynamics of Two-Qubit Gates with Motional States of Electrons on Helium}

\author{Oskar Leinonen}\email{oskar.leinonen@fys.uio.no}\author{Jonas B.~Flaten}\author{Stian D.~Bilek}\author{Øyvind S.~Schøyen}\author{Morten Hjorth-Jensen}
\affiliation{Department of Physics, University of Oslo, Oslo N-0316, Norway}

\author{Niyaz~R.~Beysengulov}
\affiliation{EeroQ Corporation, Chicago, IL 60651, USA}

\author{Zachary J. Stewart}\author{Jared D. Weidman}\author{and Angela K. Wilson}
\affiliation{Department of Chemistry, Michigan State University, East Lansing, MI 48824, USA}

\date{\today}

\begin{abstract}
Systems of individual electrons electrostatically trapped on condensed noble gas surfaces have recently attracted considerable interest as potential platforms for quantum computing. The electrons serve as charge qubits in the system, and the purity of the noble gas surface protects the relevant quantum properties of each electron. Previous work has indicated that manipulation of a confining double-well potential for electrons on superfluid helium can generate entanglement suitable for two-qubit gate operations. In this work, we incorporate a time-dependent tuning of the potential shape to further explore operation of two-qubit gates with the superfluid helium system. Through numerical time evolution of the closed system (without decoherence), we show that control-induced errors can be minimized to allow for fast, high-fidelity two-qubit gates. In particular, we simulate operation of the \sqiswap and CZ gates and obtain estimated fidelities of 0.999 and 0.996 with execution times of \SI{2.9}{\nano\second} and \SI{9.4}{\nano\second}, respectively. Furthermore, we examine the stability of these gate fidelities under non-ideal execution conditions, which reveals new properties to consider in the device design. Finally, we reflect on the impact of screening and decoherence on our results. The methodology presented here enables future efforts to isolate control-induced effects from environmental noise, which is an important step towards the realization of high-fidelity two-qubit gates with electrons on helium.
\end{abstract}

\maketitle

\section{Introduction}
Recent advances in trapping single electrons on condensed noble gas surfaces, such as liquid helium~\cite{koolstraCouplingSingleElectron2019,castoria2025} or solid neon~\cite{zhouSingleElectronsSolid2022, zhou_electron_2024,liCoherentManipulationInteracting2025,xie_2024}, have positioned these systems as promising platforms for studying light--matter interactions and for applications in quantum computing. In particular, bound electron states on liquid $^4$He surfaces offer significant potential due to their precise spatial control and shuttling capabilities~\cite{bradburyEfficientClockedElectron2011}, enabled by the defect- and impurity-free nature of the pristine helium substrate. The electron’s spin degree of freedom is predicted to exhibit long coherence times, owing to the absence of magnetic impurities in the helium dielectric environment and the extremely weak spin--orbit interaction~\cite{lyonSpinbasedQuantumComputing2006}. Additionally, both quantized out-of-plane and in-plane motional states have been proposed as qubit systems, where the spatial charge states can be tuned via applied DC potentials~\cite{platzmanQuantumComputingElectrons1999,schusterProposalManipulatingDetecting2010}. In a system of two or more electrons, quantum correlations, which are a central component in the quantum computing paradigm, can be induced by the Coulomb interaction between the charged electrons \cite{beysengulov2024,kawakami2023blueprint}. Therefore, a detailed understanding and control of the quantum degrees of freedom of both single electrons and small electron ensembles is of significant interest for quantum technologies.

The detection of individual electrons and few-electron ensembles on liquid helium has been facilitated by ongoing developments in circuit quantum electrodynamics (cQED), which serves as a foundational framework for superconducting qubit architectures and for high-sensitivity charge detection in semiconductor quantum dot systems~\cite{blaisCircuitQuantumElectrodynamics2021,burkard2023semiconductor}. Experiments have demonstrated coupling between microwave cavity photons and the cyclotron motion of a macroscopically large number of electrons on the surface of liquid helium~\cite{abdurakhimovStrongCouplingCyclotron2016}. Newer studies have shown that the motion of a single electron within an in-plane confining potential also can be coupled to the field of a microwave cavity~\cite{koolstraCouplingSingleElectron2019,castoria2025}, with strong coupling recently realized experimentally~\cite{koolstra2025strong}. Under certain conditions, the quantized electron motion can be represented as a two-level system~\cite{koolstra2025strong}. The coupled electron--cavity system can then be described by a Jaynes-Cummings-type Hamiltonian~\cite{schusterProposalManipulatingDetecting2010}, which enables the use of standard quantum non-demolition measurement techniques to infer the quantized motional state of the electron~\cite{schusterProposalManipulatingDetecting2010}. Single-qubit gates can be implemented by applying voltage drives at the qubit transition frequency via gate electrodes~\cite{blaisCircuitQuantumElectrodynamics2021}.

Several schemes have been proposed for implementing two-qubit gates, which rely on the Coulomb interaction between two electrons on the liquid helium surface~\cite{lea2000could,beysengulov2024,jennings2024quantum,kawakami2023blueprint}. Many of these proposed schemes rely on simplifications, where smooth confining potentials are typically approximated as harmonic-oscillator-type potentials near the equilibrium point. However, in general the trapping potentials are inherently nonlinear, leading to unequally spaced energy levels, and thus more complex dynamics in the system. This is especially important in developing two-qubit gates, where the electron frequencies need to be tuned by adjusting the gate electrode potentials. This tuning can cause unwanted excitations into higher excited states, limiting gate fidelities and posing challenges for achieving high-fidelity quantum operations.

The shape and placement of the gate electrodes under the helium surface are especially important because they determine how the electrons' in-plane charge states are formed and controlled. Electrodes with dimensions on the order of several hundreds of nanometers are typically used to trap a small number of electrons on the helium surface. The in-plane electrostatic confinement can then be controlled by adjusting the voltages applied to the gate electrodes~\cite{koolstraCouplingSingleElectron2019,castoria2025}. When several electrodes are involved, the trapping potential becomes an increasingly complex function of the applied voltages. We have previously studied how entanglement between two electrons can be created through the tuning of the confining potential, and how nonlinearity can be deliberately engineered and controlled to enforce specific types of correlations between the two electrons, which is a prerequisite for implementing two-qubit gates in electrons-on-helium systems~\cite{beysengulov2024}.

Here, we expand upon our earlier work by proposing two-qubit gate protocols which utilize the entanglement generation from our previous publication~\cite{beysengulov2024} and a new approach for entanglement generation. Moreover, we use a Time-Dependent Full Configuration Interaction (TD-FCI) inspired method~\cite{hochstuhl_time-dependent_2014} for distinguishable particles to simulate the dynamics of the protocol and proceed to quantify the gate performance with the average gate fidelity~\cite{press2007}. This allows us to maximize the fidelity with respect to the two parameters of the gate protocol, that is, the hold and ramp time. We specifically target the \sqiswap and CZ gates, for which we obtain fidelities of 0.999 and 0.996 with experimentally achievable execution times of \SI{2.9}{\nano\second} and \SI{9.4}{\nano\second}, respectively. Finally, we examine the sensitivity of these gate fidelities to perturbations in the hold and ramp time. Our results can aid the experimental realization of two-qubit gates with electrons-on-helium systems.

The paper is arranged as follows: Section~\ref{sec:device} describes the device used in our simulations, and the operation of the proposed gate protocol is detailed in Sec.~\ref{sec:gate-operation}. The numerical method used to optimize the parameters of the protocol is outlined in Sec.~\ref{sec:num-method}. In Sec.~\ref{sec:res-swap} and Sec.~\ref{sec:res-cz}, we present the results for the \sqiswap and CZ gates, respectively. The sensitivity of the gates to ramp and hold time deviations is analyzed in Sec.~\ref{sec:sensitivity} before we present our conclusions and future outlooks in Sec.~\ref{sec:conclusions}. Some more in depth analyses are provided in the appendices.

\section{Device and Operation}
\label{sec:method}
Electrons near the surface of liquid helium polarize the helium atoms, resulting in a weak attractive force that pulls them toward the surface. At the helium--vacuum interface, the electrons experience hard-core repulsion with a potential barrier of around \SI{1}{\electronvolt}, which prevents them from penetrating into the liquid. The resulting out-of-plane interaction with the helium surface leads to a quantized Rydberg-like energy spectrum, and at sufficiently low temperatures $T < \SI{2}{\kelvin}$, the electrons only occupy the ground state. The electron’s in-plane motion is controlled by a confining potential created by electrodes positioned around and beneath the helium surface. This work builds upon our previous publication~\cite{beysengulov2024}, utilizing the same device geometry and entanglement generation scheme reported in that study. Here, we expand our analysis of the system by proposing two-qubit gate protocols and simulating the dynamics of the system, in order to predict and optimize the fidelities of the two-qubit gates.

\subsection{Device description}
\label{sec:device}
\begin{figure}[ht]
    \centering
    \includegraphics[width=\linewidth]{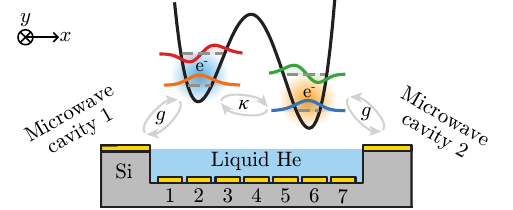}
    \caption{Schematic microdevice, in which two electrons are trapped on the surface of a liquid helium basin in an electrostatic double-well potential created by electrodes 1–7. The two-qubit gates are driven by the Coulomb interaction $\kappa$ between the two electrons, while single qubit gates and readout are performed through the electron--cavity interaction $g$.}
    \label{fig:device}
\end{figure}

The device geometry is schematically depicted in Fig.~\ref{fig:device}. It consists of seven control gate electrodes designed to create an electrostatic double-well potential on the helium surface, which confines two electrons in separate potential wells. The electrodes are \SI{200}{\nano\meter} wide, spaced by \SI{200}{\nano\meter} and positioned beneath a \SI{500}{\nano\meter} deep layer of liquid helium. This geometry was chosen to create an in-plane motional quantization axis along the $x$ direction, with energy gaps in the frequency range of 5–\SI{15}{\giga\hertz}. These states are decoupled from motional states along the $y$ direction at approximately six times larger frequencies, thereby allowing us to ignore the states along the $y$ axis~\cite{beysengulov2024}. Each electron is dipole-coupled to its own microwave cavity, enabling both qubit state readout and operation of single-qubit gates. The Coulomb interaction between the two electrons, which are separated by a distance of approximately \SI{1.5}{\micro\meter} for the well configurations used in this study, provides a mechanism through which we can introduce the necessary entanglement to operate two-qubit gates.

The Hamiltonian of the system is given by
\begin{align}
        \hat{H}
        = \sum_{i = 1}^{2}\qty(
            -\frac{1}{2}\dv[2]{}{x_i}
            + v(x_i)
        )
        + u(x_1, x_2),
        \label{eq:hamiltonian}
\end{align}
where $u(x_1, x_2)=\kappa/\sqrt{(x_1-x_2)^2+\epsilon^2}$ is the Coulomb interaction with a strength parameter $\kappa=2326$ determined by the choice of energy unit \cite{beysengulov2024}, and $\epsilon=0.01$ is a shielding parameter introduced to remove the singularity at $x_1=x_2$ \cite{kvaal2007}. Due to the proximity of the underlying electrodes, the Coulomb interaction is reduced by screening effects. However, we employ an unscreened interaction in our analysis, which provides an upper bound for the interaction strength. We discuss the implications of the screening in Section~\ref{sec:screening}. The electrostatic potential from the electrodes beneath the condensed helium layer is given by
\begin{equation}
    v(x) = -\sum_{k=1}^7\alpha_k(x)V_k \equiv -\vb*{\alpha}(x)\cdot\vb*{V},
    \label{eq:potential}
\end{equation}
where $V_k$ is the voltage applied to electrode $k$ and $\alpha_k(x)$ is a dimensionless function quantifying the relative contribution of the electrode $k$ to the electrostatic potential at the helium surface, which is calculated using the Finite Element Method (FEM)~\cite{beysengulov2024}.

\subsection{Gate operation}
\label{sec:gate-operation}
In order to introduce dynamics to the system, we parametrize the voltage vector in Eq.~\eqref{eq:potential} through \textit{voltage functions} $\vb*{V}(\lambda)$. These functions are constructed such that the system can alternate between different \textit{configurations}, i.e, shapes of the electrostatic double-well potential, each with distinct properties. Following Ref.~\cite{beysengulov2024} we distinguish between three configurations: In configuration~I the system is in an idle state, where the effective interaction leading to entanglement between the two electrons is minimized, thus enabling readout of the qubit states as well as single-qubit gate operation. Configuration~II is designed to favor an interaction suitable for SWAP-type gates, whereas the interaction in configuration~III is tailored to facilitate for Controlled-Z (CZ) gates. Incorporating the voltage vectors of such configurations, two different voltage functions will be explored in this work. The system properties employing these functions are presented in Ref.~\cite{beysengulov2024} and Appendix~\ref{app:zz-opt}, respectively. The first voltage function, which we will denote $\vb*{V}^\beta$, reads 
\begin{equation}
\label{eq:Vbeta}
    \vb*{V}^\beta(\lambda) = (1-\lambda)\vb*{V}^\beta_\mathrm{I} + \lambda\vb*{V}^\beta_\mathrm{III},
\end{equation} where $\vb*{V}^\beta_\mathrm{I}$ and $\vb*{V}^\beta_\mathrm{III}$ are the constant voltage vectors from our previous work, which were optimized for opposite-sign anharmonicities $\beta^L = - \beta^R$ in the two wells \cite{beysengulov2024, zhao2020}. Opposite-sign anharmonicity was however shown to be insufficient to suppress unwanted ZZ-coupling in the entanglement driven by the Coulomb interaction \cite{beysengulov2024}. Therefore, we also present results for a second voltage function, denoted $\vb*{V}^\zeta$, given by
\begin{equation}
\label{eq:Vzeta}
   \vb*{V}^\zeta(\lambda) = 
    (1-\lambda)\vb*{V}^\zeta_\mathrm{I} + \lambda\vb*{V}^\zeta_\mathrm{II/III},
\end{equation} 
where the last term is set to either $\vb*{V}^\zeta_\mathrm{II}$ or $\vb*{V}^\zeta_\mathrm{III}$, depending on the desired type of entanglement. For this voltage function, the main goal in the optimization of the constant voltage vectors $\vb*{V}^\zeta_\mathrm{I}$, $\vb*{V}^\zeta_\mathrm{II}$ and $\vb*{V}^\zeta_\mathrm{III}$ was to directly minimize the ZZ-coupling, quantified by
\begin{equation}
\zeta = E_4 - E_2 - E_1 + E_0,
\end{equation}
where $E_i$ is the $i$-th eigenenergy of the full system. This optimization is described in more detail in appendix~\ref{app:zz-opt}, together with the resulting energy spectra and von Neumann entropies.

Two-qubit gate operations are enabled by introducing a time-dependence in the parametrization of the voltage functions, $\vb*{V}(\lambda)=\vb*{V}(\lambda(t))$. The time-dependent parameter $\lambda(t)$ should induce a smooth transition from the idle configuration~I to one of the entangling configurations (II or III) and then back, as time progresses \cite{strauch_quantum_2003}. Two-qubit gates can then be realized by driving the system into an entangling configuration, where the system is kept for a duration $\thd$ to generate entanglement, before returning to the configuration~I, where the electronic states can be regarded as qubit states. However, at intermediate timescales, where the transition between the configurations is neither fully diabatic nor adiabatic, entanglement is also generated during the transition itself. To accurately capture the full dynamics of the gate protocol, we employ a numerical method, described in section~\ref{sec:num-method}, which provides an exact solution to the time-dependent problem across all relevant timescales. 

Inspired by~\cite{ghosh2013}, the explicit function shape of the time-dependent parameter $\lambda(t)$ was chosen to be
\begin{equation}
\mathtoolsset{multlined-width=0.9\displaywidth}
\begin{multlined}
    \lambda(t) = \frac{\lambda\sbt{max}}{2}\Bigg[\erf\left(\frac{t-\frac{1}{2}t_\mathrm{ramp}}{\sqrt{2}\sigma}\right) \\ 
    - \erf\left(\frac{t-t_\mathrm{gate}+\frac{1}{2}t_\mathrm{ramp}}{\sqrt{2}\sigma}\right)\Bigg],
    \label{eq:lambdat}
\end{multlined}
\end{equation}
where $\lambda\sbt{max}$ is the maximum value the function assumes, $\tgt = \thd + 2\trp$ is the total gate time, $\thd$ is the time for which the system is kept in an entangling configuration (II or III) and $\trp$ is how quickly the system transitions between the idle configuration~I and the desired entangling configuration. The function is illustrated in Fig.~\ref{fig:ramp_func}. Throughout this work, we let $\trp = 4\sqrt{2}\sigma$ \cite{ghosh2013}.

For both voltage functions, configuration~I is located at $\lambda(t) = 0$, i.e. when $t=0$ or $t>\tgt$, while the entangling configurations are realized when $t \in (\trp, \trp + \thd)$. For the first voltage function, \vbeta, the type of entanglement, i.e. configuration~II or III, is decided by setting $\lambda\sbt{max}$ to $\lambda_\mathrm{II} \equiv 0.46$ or $1$, respectively \cite{beysengulov2024}. For the second voltage function, \vzeta, $\lambda\sbt{max}$ should be $1$ for both types of entanglement, and the configuration is instead determined by choosing either $\vb*{V}^\zeta_\mathrm{II}$ or $\vb*{V}^\zeta_\mathrm{III}$ in the second term of Eq.~\eqref{eq:Vzeta}. In order to utilize the entanglement of configurations~II and III to drive specific two-qubit gates, what remains is then to determine the hold and ramp times that give the best gate performance. 

\subsection{Optimization of gate fidelity}
\label{sec:num-method}
\begin{figure}[ht]
    \centering
    \includegraphics[width=\columnwidth]{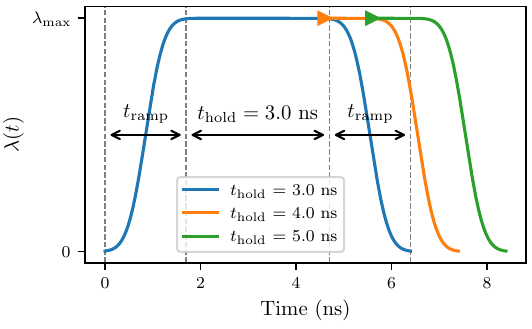}
    \caption{The blue line shows the shape of $\lambda(t)$ with $\trp = \SI{0.3\times4\sqrt{2}}{\nano\second}\approx \SI{1.7}{\nano\second}$ and $\thd = \SI{3.0}{\nano\second}$. The orange and green lines illustrate how calculations with longer hold times, in this example \SI{4.0}{\nano\second} and \SI{5.0}{\nano\second} respectively, can be warm-started using previous calculations, thus reducing the computational cost.}
    \label{fig:ramp_func}
\end{figure}

We employ a grid search approach to optimize the ramp and hold times for maximum gate fidelity. The efficiency of the search is improved by noting that for each fixed value of the ramp time, the calculation of a new hold time can be warm-started using the previous calculation, as illustrated in Fig.~\ref{fig:ramp_func}.

In our numerical method, we assume that the potential wells are deep enough for the two trapped electrons to effectively become distinguishable \cite{beysengulov2024}. This assumption allows us to expand any two-body state $\ket{\Psi(t)}$ in a tensor product basis of sinc-DVR \cite{colbert1992} functions, 
\begin{equation}
    \ket{\Psi(t)} = \sum_{\alpha\beta}C_{\alpha\beta}(t)\ket{\chi\spL_\alpha\chi\spR_\beta},
    \label{eq:Psi}
\end{equation}
 where $\chi\spL_\alpha$ and $\chi\spR_\beta$ represent two distinct sets of sinc-DVR functions defined on discrete grid points in the left and right well, respectively. Using the matrix elements derived in appendix B of~\cite{beysengulov2024}, we can then write down the effect of $\hat{H}(t)$ on a state $\ket{\Psi(t)}$ as 
\begin{equation}
    \mathtoolsset{multlined-width=0.9\displaywidth}
    \begin{multlined}
\left(\hat{H}(t)\ket{\Psi(t)}\right)_{\gamma\delta} \equiv \bra{\chi\spL_\gamma\chi\spR_\delta}\hat{H}(t)\ket{\Psi(t)} =\\
\sum_{\alpha\beta}C_{\alpha\beta}(t)\Big[\delta_{\delta\beta}\Big(t\spL_{\gamma\alpha} + v\spL_\alpha\big(\lambda(t)\big)\Big) \\
+ \delta_{\gamma\alpha}\Big(t\spR_{\delta\beta} + v\spR_\beta\big(\lambda(t)\big)\Big) + \delta_{\gamma\alpha}\delta_{\delta\beta}u_{\alpha\beta}\Big],
    \end{multlined}
    \label{eq:Hc}
\end{equation}
where the subscripts on the left-hand side denote the coefficient matrix of the state $\hat{H}(t)\ket{\Psi(t)}$ for a tensor product expansion like~\eqref{eq:Psi}. With this expression established, the lowest energy eigenstates can be determined iteratively using the Davidson method~\cite{davidson1975,sun2020}. We compute the six lowest eigenstates at $t=0$, i.e. when the system is in configuration~I. These eigenstates correspond to the computational basis states $\ket{00}, \ket{01}, \ket{10}, \ket{02}, \ket{11}$ and $\ket{20}$, respectively \cite{beysengulov2024}. Among them, the four relevant qubit states $\ket{00}, \ket{01}, \ket{10}$ and $\ket{11}$, i.e. the eigenstates $\ket{\Phi_0}, \ket{\Phi_1}, \ket{\Phi_2}$ and $\ket{\Phi_4}$, are propagated in time using the Crank-Nicolson propagator~\cite{press2007}. We denote the propagated state which initially is equal to eigenstate $\ket{\Phi_n}$ by $\ket{\Psi_n(t)}$. The system is evolved for a time $t_\mathrm{gate} = t_\mathrm{hold} + 2t_\mathrm{ramp}$ with an integration step size of $\Delta t =(\omega_{\mathrm{qubit}}/2\pi)^{-1}/100 = \SI{0.001}{\nano\second}$, suitable for dynamics on the qubit energy scale of $\omega_{\mathrm{qubit}}/2\pi \sim \SI{10}{\giga\hertz}$, which is typical for our system~\cite{beysengulov2024}. After the time evolution, the $4\times 4$ matrix $U$ describing the qubit dynamics is given by the overlap matrix between the four two-qubit states at $t=0$ and $t=\tgt$, 
\begin{equation}
    U_{ij} = \bra{\Phi_{n_i}}\ket{\Psi_{n_j}(t_\mathrm{gate})},
\label{eq:Uij}
\end{equation}
where $n_k = k$ for $k \in \{0, 1, 2\}$ while $n_3 = 4$ (omitting state 3 from the matrix since it corresponds to the unwanted $\ket{02}$ state). Single-qubit rotations $R(\theta\sbL, \theta\sbR) = R_z(\theta\sbL)\otimes R_z(\theta\sbR)$ \cite{nielsen_quantum_2010} are then applied, yielding an overall two-qubit gate matrix $G(\theta\sbL, \theta\sbR) = R(\theta\sbL, \theta\sbR)U$.
 
 We quantify the performance of the gate $G(\theta\sbL, \theta\sbR)$ by the average gate fidelity \cite{pedersen2007}
 \begin{equation}
     F = \left(\Tr( MM^\dagger) + \abs{\Tr(M)}^2\right)/20.
     \label{eq:fid}
 \end{equation}
 Here, $M \equiv U_0^\dagger G(\theta\sbL, \theta\sbR)$, where $U_0$ is the gate matrix describing an ideal operation of the targeted two-qubit gate. The single-qubit rotation angles $\theta\sbL^\text{opt}$ and $\theta\sbR^\text{opt}$ that maximize the fidelity are found using a standard optimizer~\cite{2020SciPy-NMeth}, and the final optimized two-qubit gate is then given by $G(\theta\sbL\spt{opt}, \theta\sbR\spt{opt})$.

\section{Results and Discussion}
In this section we present results from the simulation of two different two-qubit gates, the \sqiswap gate and the CZ gate, for both of the voltage functions described in section~\ref{sec:gate-operation}. The gate accuracy up to single-qubit rotations is quantified using the fidelity measure from Eq.~\eqref{eq:fid} and the gate time is reported. 

\subsection{\texorpdfstring{\sqiswap}{√iSWAP} gate}
\label{sec:res-swap}
\begin{figure}[ht!]
    \centering
    \includegraphics[width=\linewidth]{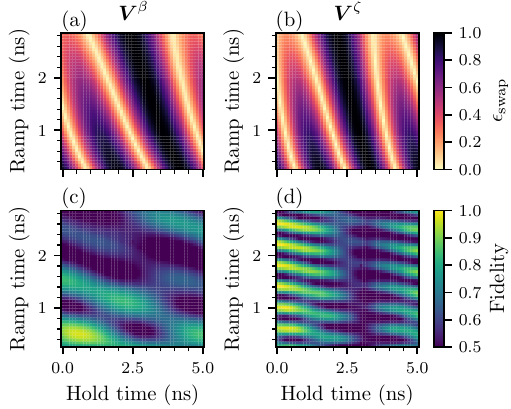}
    \caption{Grid search for a \sqiswap gate with step sizes $\Delta\trp = \SI{0.01\times4\sqrt{2}}{\nano\second} \approx \SI{0.06}{\nano\second}$ and $\Delta t_\mathrm{hold} = \SI{0.1}{\nano\second}$. The plots in the top row show the swap error $\esp$ for \vbeta in~(a) and \vzeta in~(b). The corresponding gate fidelities of the two voltage functions are displayed in plots~(c) and (d), respectively. The maximum fidelity achieved for \vbeta is $F=0.971$, obtained at $\trp=\SI{0.51}{\nano\second}$ and $\thd=\SI{0.6}{\nano\second}$, and for \vzeta the maximum fidelity is $F=0.999$, realized with $\trp=\SI{1.41}{\nano\second}$ and $\thd=\SI{0.1}{\nano\second}$.}
    \label{fig:iswap-fid}
\end{figure}

The first two-qubit gate we strive to achieve is the \sqiswap gate,
\begin{equation}
    \sqrt{i\mathrm{SWAP}} = \begin{pmatrix}
        1 & 0 & 0 & 0\\
        0 & 1/\sqrt{2} & i/\sqrt{2} & 0\\
        0 & i/\sqrt{2} & 1/\sqrt{2} & 0\\
        0 & 0 & 0 & 1
    \end{pmatrix}.
    \label{eq:sqrtiswap}
\end{equation}
For a perfect \sqiswap gate, a pure $\ket{01}$ or $\ket{10}$ state should be swapped into a maximally entangled state with equal probability of $\ket{01}$ and $\ket{10}$. We can thus define a \textit{swap error} to measure the deviation from the ideal probabilities, where 0 indicates perfect agreement and 1 represents the maximum error, according to
\begin{equation}
    \esp = \sum_{i,j=1}^2 \abs{0.5 - \abs{U_{ij}}^2}/2,
\end{equation}
where $U_{ij}$ is given by Eq.~\eqref{eq:Uij}. We conduct a sweep over hold times from 0~to~\SI{5}{\nano\second} with steps of $\Delta\thd = \SI{0.1}{\nano\second}$, and over ramp times from $\trp=\SI{0.05\times4\sqrt{2}}{\nano\second}\approx\SI{0.28}{\nano\second}$ to $\trp=\SI{0.5\times4\sqrt{2}}{\nano\second}\approx\SI{2.83}{\nano\second}$, with steps of $\Delta\trp = \SI{0.01\times4\sqrt{2}}{\nano\second}\approx\SI{0.06}{\nano\second}$. The obtained swap error $\esp$ for the two different voltage functions is depicted in Fig.~\ref{fig:iswap-fid}~(a) for \vbeta and Fig.~\ref{fig:iswap-fid}~(b) for \vzeta, with the corresponding gate fidelities shown in Fig.~\ref{fig:iswap-fid}~(c)~and~(d), respectively.

Comparing the four plots in Fig.~\ref{fig:iswap-fid}, we find that $\esp$ behaves similarly across the two voltage functions. However, identifying a meaningful resemblance between $\esp$ and the fidelity is challenging, demonstrating that $\esp$ is a bad predictor of the gate fidelity and thus that the relative phases of the states could be more important than their amplitudes. The structure of the fidelity differs substantially between the two voltage functions, with the fidelity decaying significantly with increased ramp time for \vbeta. The reason for this decay is mainly the poor suppression of the ZZ-coupling provided by the opposite-sign anharmonicities \cite{beysengulov2024}. Both cases reveal oscillatory behavior as a function of ramp time. This oscillation is linked to the energy difference between the first and second excited states of the system, which in turn establishes the phase difference of their respective matrix elements. This phase difference is crucial for the performance of the \sqiswap gate. A more detailed analysis of the features in the fidelity plots is provided in appendix~\ref{app:gate_details}.

The maximum fidelity achieved by voltage function \vbeta is $F=0.971$, with $\trp = \SI{0.51}{\nano\second}$ and $t_\mathrm{hold} = \SI{0.6}{\nano\second}$, corresponding to a total gate time of \SI{1.6}{\nano\second}. A more accurate gate is acquired with voltage function \vzeta, reaching an optimal fidelity of $F=0.999$ for $\trp = \SI{1.41}{\nano\second}$ and $\thd = \SI{0.1}{\nano\second}$, corresponding to a total gate time of \SI{2.9}{\nano\second}. This gate operates more slowly than the gate associated with \vbeta; however, it should be noted that if the total execution time is limited to that of the \vbeta gate (durations around \SI{1.6}{\nano\second}), both voltage functions achieve the same fidelity.

The matrices of the best gates we obtained can be written as
\begin{equation*}
    G^\beta_{\sqrt{i\mathrm{SWAP}}} = 
    \begin{pmatrix}1.00 & 0 & 0 & 0 \\
    0 & \sqrt{0.51}e^{ i0.12 \pi} & \sqrt{0.49}e^{ i0.61 \pi} & 0 \\
    0 & \sqrt{0.49}e^{ i0.63 \pi} & \sqrt{0.51}e^{ i0.11 \pi} & 0 \\
    0 & 0 & 0 & 1.00 
    \end{pmatrix}
\end{equation*}
for \vbeta and 
\begin{equation*}
   G^\zeta_{\sqrt{i\mathrm{SWAP}}} = \begin{pmatrix}
1.00 & 0 & 0 & 0\\ 
0 & \sqrt{0.53} & \sqrt{0.47}e^{ i0.49 \pi} & 0 \\
0 & \sqrt{0.47}e^{ i0.50 \pi} & \sqrt{0.53} & 0 \\
0 & 0 & 0 & 1.00
 \end{pmatrix}
\end{equation*}
for \vzeta. Despite the higher fidelity displayed by the \vzeta gate, it is evident that the amplitudes of the $\ket{01}\leftrightarrow\ket{10}$ swap are worse than for the \vbeta gate, further strengthening the claim that the relative phases are the most important factors for the \sqiswap gate fidelity. Nevertheless, the gate fidelity is influenced by both the amplitude and phase of the elements associated with the swap, as well as the amplitude and phase of the $G_{33}$ element. As the ZZ-coupling decreases, the variation in the latter element reduces, thus increasing the probability that all matrix elements obtain desirable values for a certain combination of $\thd$ and $\trp$, which is why the \vzeta gate achieves better fidelity.

\subsection{CZ gate}
\label{sec:res-cz}
\begin{figure}[ht]
    \centering
    \includegraphics[width=\linewidth]{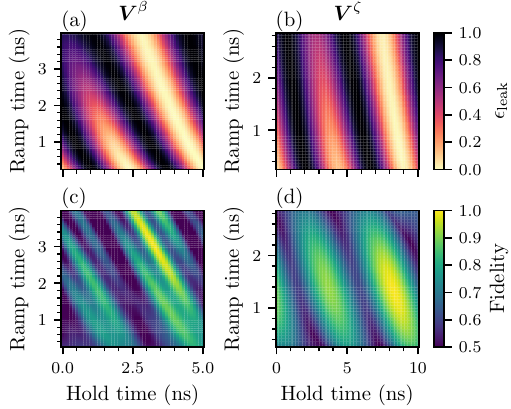}
    \caption{Grid search for a CZ gate with step sizes $\Delta\trp = \SI{0.01\times4\sqrt{2}}{\nano\second}\approx\SI{0.06}{\nano\second}$ and $\Delta\thd = \SI{0.1}{\nano\second}$. Note that the ranges of the axes are different between the two voltage functions. The plots in the top row show the leakage error $\elk$ for \vbeta in (a) and \vzeta in (b). The corresponding gate fidelities are displayed in plot~(c) and (d), respectively. The maximum achieved fidelity is $F=0.996$ for both voltage functions. With \vbeta this value is obtained at $\trp=\SI{3.11}{\nano\second}$ and $\thd=\SI{3.2}{\nano\second}$, whereas it is given by $\trp=\SI{1.30}{\nano\second}$ and $\thd=\SI{8.3}{\nano\second}$ with \vzeta.}
    \label{fig:cz-fid}
\end{figure}

The search for a high-fidelity CZ gate is performed in a similar fashion to the case of the $\sqrt{i\mathrm{SWAP}}$ gate described in the previous section, but with the voltage functions modified to arrive at configuration~III after time $t=\trp$, as described in section~\ref{sec:gate-operation}. Expressed as a matrix, the CZ gate reads
\begin{equation}
    \mathrm{CZ} = \begin{pmatrix}
        1 & 0 & 0 & 0\\
        0 & 1 & 0 & 0\\
        0 & 0 & 1 & 0\\
        0 & 0 & 0 & -1
    \end{pmatrix}.
    \label{eq:CZ}
\end{equation}
Thus, the gate produces a $\pi$ phase shift on the $\ket{11}$ state while preserving all other states. We aim to achieve this phase shift by strengthening the interactions among the $\ket{02}, \ket{11}$ and $\ket{20}$ states. However, this approach entails a risk of population leakage from $\ket{11}$ to the non-qubit states $\ket{02}$ and $\ket{20}$. We therefore define a \textit{leakage error}
\begin{equation}
    \elk = 1- \abs{U_{33}}^2,
\end{equation}
where $U_{33} = \bra{\Phi_{4}}\ket{\Psi_{4}(t_\mathrm{gate})}$, see Eq.~\eqref{eq:Uij}. We then perform a sweep over $\thd$ and $\trp$, utilizing the same step sizes as in the \sqiswap case, but with the search area increased such that the ramp time ranges from $\trp=\SI{0.05\times4\sqrt{2}}{\nano\second}\approx\SI{0.28}{\nano\second}$ to $\trp=\SI{0.7\times4\sqrt{2}}{\nano\second}\approx\SI{3.96}{\nano\second}$ for \vbeta and the hold time ranges from 0~to~\SI{10}{\nano\second} for \vzeta. Fig.~\ref{fig:cz-fid} visualizes the results from the sweep, with the leakage error $\elk$ shown in plot (a) for \vbeta and plot (b) for \vzeta. The corresponding gate fidelities are depicted in Fig.~\ref{fig:cz-fid}~(c)~and~(d), respectively. 

In contrast to the amplitude based swap error of the \sqiswap gate, we observe that the leakage error $\elk$, which is also based solely on the amplitudes, matches the CZ gate fidelity quite well. Some additional structure emerges for the \vbeta fidelity, shown in Fig.~\ref{fig:cz-fid}~(c), due to unwanted interactions between the $\ket{01}$ and $\ket{10}$ states during the transition from configuration~I to configuration~III. This interaction reduces the fidelity unless the hold time compensates for the oscillations between these states, as detailed in appendix~\ref{app:gate_details}. The other key difference between the two voltage functions is the oscillation frequency of both the fidelity and $\elk$ relative to hold time, which is about twice as large for \vbeta. (Note the different ranges on the x-axes in Fig.~\ref{fig:cz-fid}.) This frequency results from the dynamics among the three excited states $\ket{\Psi_3(t)}, \ket{\Psi_4(t)}$ and $\ket{\Psi_5(t)}$. The same dynamics are also responsible for the fidelity not attaining its maximum until the second period of the oscillation, and only at certain ramp times. However, the intricacies of the time evolution make it challenging to directly attribute the behavior of these dynamics to specific characteristics of the different voltage functions.

The best gate fidelity achieved with voltage function \vbeta is $F=0.996$, which is obtained for $\trp = \SI{3.11}{\nano\second}$ and $\thd = \SI{3.2}{\nano\second}$, yielding a total gate time of \SI{9.4}{\nano\second}. Voltage function \vzeta produces the same fidelity, $F=0.996$, however it is obtained at $\trp = \SI{1.30}{\nano\second}$ and $\thd = \SI{8.3}{\nano\second}$, resulting in a slower execution time of \SI{10.9}{\nano\second}. The reduced speed is a result of the lower frequency discussed above. In terms of matrices, the best gate we achieve with \vbeta can be expressed as\footnote{The two center column vectors only appear to not be normalized because the values are rounded.} 
\begin{equation*}
    G^\beta_\mathrm{CZ} = 
    \begin{pmatrix}
    1.00 & 0 & 0 & 0 \\
    0 & 1.00e^{ -i0.01 \pi} & 0.02e^{ -i0.29 \pi} & 0 \\
    0 & 0.02e^{ -i0.70 \pi} & 1.00e^{ -i0.01 \pi} & 0 \\
    0 & 0 & 0 & 0.99e^{ -i1.00 \pi} 
    \end{pmatrix},
\end{equation*}
and with \vzeta can be written as\footnotemark[1]
\begin{equation*}
    G^\zeta_\mathrm{CZ} =
    \begin{pmatrix}
    1.00 & 0 & 0 & 0 \\
    0 & 1.00 & 0.02e^{ i0.67 \pi} & 0 \\
    0 & 0.03e^{ i0.56 \pi} & 1.00 & 0 \\
    0 & 0 & 0 & 0.99e^{ -i1.00 \pi}
    \end{pmatrix}.
\end{equation*}
We note that both gate matrices have some unwanted off-diagonal elements, as well as a minor leakage of the $G_{33}$ element. Additionally, \vbeta has a small phase error on the diagonal $G_{11}$ and $G_{22}$ elements. Still, all these errors are small, hence the high fidelities. 

\begin{figure*}[ht!]
    \centering
    \includegraphics[width=\linewidth]{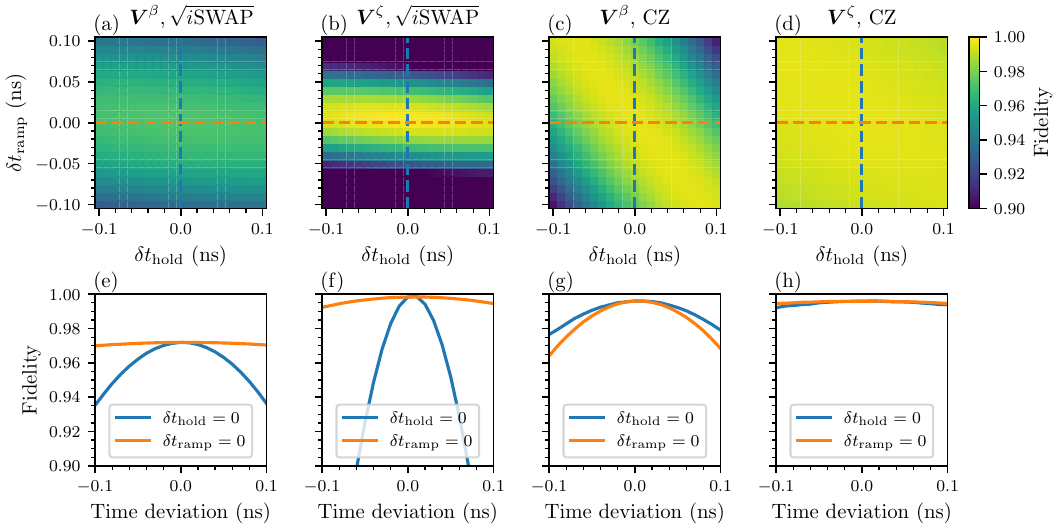}
    \caption{Gate fidelity for small deviations $\delta\trp$ and $\delta\thd$ from the optimal ramp and hold time. The change in fidelity is presented for the \sqiswap gate in plot~(a) using \vbeta, and in plot~(b) using \vzeta. For the CZ gate, the same results are shown in plot~(c) with \vbeta and in plot~(d) with \vzeta. Corresponding one-dimensional cross-sections are presented in plots~(e)--(h). Note that the ramp time is crucial for the \sqiswap gate's performance.}
    \label{fig:error}
\end{figure*}

\subsection{Sensitivity to ramp and hold deviations}
\label{sec:sensitivity}
Finally, we conduct a more thorough investigation into the stability of the gate fidelity as the hold and ramp times deviate from the ideal values identified in the previous two sections. The fidelities of the gates in a window of $\pm\SI{0.1}{\nano\second}$ around the optimal values of $\thd$ and $\trp$ are presented in plots~(a)--(d) of Fig.~\ref{fig:error}, with corresponding one-dimensional cross-sections shown in plots~(e)--(h). In these calculations, the resolution is \SI{0.01}{\nano\second} for both hold and ramp times.

In the case of the \sqiswap gate, shown in Fig.~\ref{fig:error} for \vbeta ((a) and (e)) and for \vzeta ((b) and (f)), the gate performance remains relatively stable with non-ideal hold times. However, it is very sensitive to deviations in ramp time, particularly for the \vzeta voltage function. These properties are a result of the phase obtained through the $\ket{01}\leftrightarrow\ket{10}$ swap (see appendix~\ref{app:gate_details} for details). During the ramp up from configuration~I to configuration~II, the average energy difference between the $\ket{\Psi_1(t)}$ and $\ket{\Psi_2(t)}$ states is \SI{0.98}{\giga\hertz} for \vbeta and \SI{2.52}{\giga\hertz} for \vzeta. In contrast, the same energy difference during the hold and ramp down stages of the gate operation is only about \SI{0.1}{\giga\hertz}. The accumulated phase differences of the corresponding matrix elements are determined by integrals of the energy difference over time. Consequently, the relatively large energy difference during the ramp-up stage causes the phase difference to predominantly be determined by the ramp time. This explains the observed sensitivity of gate fidelity with respect to ramp time deviations, while the relatively small energy difference during the hold stage explains the stability to deviations in hold time.

The smaller energy difference, and thus better stability, of the gate associated with \vbeta can be attributed to two factors. Primarily, the energy difference in configuration~I is smaller, and secondly, the energies of the states cross during the time evolution, resulting in $\ket{\Psi_1(t)}$ having a higher energy than $\ket{\Psi_2(t)}$ after the ramp up is finished, thereby reducing the average energy difference. Both these factors should be incorporated in future optimizations of electrode voltages $\vb*{V}_\mathrm{II}$ in order to improve the stability of the \sqiswap gate.

The CZ gate with the \vbeta voltage function exhibits about the same sensitivity for both parameters, see Fig.~\ref{fig:error} (c) and (g), with a slightly larger dependence on the hold time. Similarly to before, this sensitivity mainly arises from interactions between the $\ket{\Psi_1(t)}$ and $\ket{\Psi_2(t)}$ states, which are entirely unwanted for this gate. However, in this case the energy difference between these states is roughly the same during all three stages of the gate execution, thus explaining the similar sensitivity with respect to both $\trp$ and $\thd$.

All three gates discussed thus far might be problematic to implement experimentally due to the relatively fast drop in fidelity if the precision by which the gate parameters can be tuned is insufficient. The last gate, CZ implemented with the \vzeta voltage function, is on the contrary highly stable, mainly because interactions between $\ket{\Psi_1(t)}$ and $\ket{\Psi_2(t)}$ remain suppressed throughout the whole gate protocol. As seen in Fig.~\ref{fig:error}~(d) and (h), the drop in fidelity is less than one percent for ramp and hold time deviations of \SI{0.1}{\nano\second}, which should allow decent gate performance with current control electronics~\cite{ding2024experimental}.

\subsection{Screening and Decoherence}
\label{sec:screening}
We conclude our discussion by addressing two physical effects omitted from our current model that would impact the performance of an experimental realization: screening of the Coulomb interaction between the electrons and decoherence of the electron motional states.

The proximity of the electrons to the metallic electrodes beneath the helium layer gives rise to a screening of the Coulomb interaction by free charges in the electrodes. To estimate the impact of this screening on the two-qubit gates, it is crucial to isolate its effect on the two-particle coupling strength that leads to entanglement.

As detailed in Appendix~\ref{app:screening}, we model the screening using the method of image charges. An expansion of the interaction potential reveals that the screening has two distinct contributions. The constant and first-order terms in the expansion lead to shifts in the total energy and equilibrium positions. While significant, these shifts primarily affect the qubit frequencies and the resonance conditions for the gates. In our protocol, these effects can be compensated for by modifying the external potential through re-optimization of the voltage functions, $\vb*{V}(\lambda)$, to restore the necessary avoided crossings between energy levels.

The second contribution arises from the curvature of the screened interaction potential, which couples the displacements of the two electrons. It is this entangling coupling that drives the gate dynamics and determines the gate speed. In Appendix~\ref{app:screening}, we derive the screening factor $\eta$ for this coupling strength. For our device geometry, with an electron separation of $d\approx\SI{1.5}{\micro\meter}$ and helium depth $h=\SI{500}{\nano\meter}$, we calculate a screening factor of $\eta \approx 0.69$, which implies an increase in gate time by a factor of roughly $1/\eta \approx 1.45$. This moderate increase in gate time suggests that the high fidelities reported in this work can be maintained even in the presence of screening, provided the above-mentioned retuning of the voltage functions is implemented. 

Furthermore, the overall reduction of the Coulomb interaction strength due to screening is beneficial for our voltage optimization, as it effectively localizes the regions in voltage space where different motional states of the electrons couple. This localization improves the isolation of desired interactions from unwanted ones in the different configurations. Consequently, the fidelities reported in this work, where decoherence effects are not considered, would increase.

In a realistic experimental scenario, the performance of the two-qubit gates will, however, be limited by decoherence of the electron motional states. Recent experiments~\cite{koolstra2025strong} indicate that the charge qubit lifetime is predominantly limited by dephasing, with measured rates on the order of \SI{60}{\mega\hertz}. This is comparable to the two-qubit gate rates obtained in this work, implying that further suppression of dephasing will be necessary to achieve high-fidelity gates. Although the underlying source of this dephasing is not yet fully established, two primary mechanisms have been proposed: fluctuations from background charges localized on the helium surface near the electron, and coupling to a bath of thermally excited ripplons (surface capillary waves). In the former case, the fluctuating charges effectively modulate the electrostatic potential, causing random variations in the qubit frequency resulting in motional dephasing. This mechanism may be eliminated through device and cell design improvements aimed towards more control of the generation and trapping of electrons on the helium surface. 

Regarding the latter mechanism, the presence of thermally excited ripplons modulates the energies of the electron motional states, which leads to dephasing. (See Appendix~\ref{sec:ehe-ripplon} for an analysis of the coupling Hamiltonian.) The electron coupling to the ripplonic bath can be quantified by a dimensionless parameter $\mathcal{C}$, following Ref.~\cite{dykmanSpinDynamicsQuantum2023}. By varying the external pressing field $E_{\perp}$, the coupling parameter can be tuned from a strong regime $\mathcal{C} \gg 1$ to a weak regime $\mathcal{C} \ll 1$. In the strong coupling regime the electron absorption spectrum has a Gaussian shape with a characteristic width $\gamma \propto \sqrt{\mathcal{C}}$~\cite{dykman2025}. For temperatures $T = \SI{10}{\milli\kelvin}$ and pressing fields $E_{\perp} = \SI{1000}{\volt\per\centi\meter}$, the parameter $\mathcal{C}$ is about $10$ and the coupling to ripplons is strong. This gives a linewidth of the absorption spectrum of $\gamma/2\pi \approx \SI{100}{\mega\hertz}$, and in this regime coherent Rabi oscillations between electron motional states are not expected. Lower pressing fields are required to get to the weak coupling regime. For $E_{\perp} = \SI{10}{\volt\per\centi\meter}$ the linewidth is $\gamma/2\pi \approx \SI{10}{\mega\hertz}$, however in this regime a narrow zero-ripplon line (ZRL) appears atop the broader Gaussian background. This is analogous to the zero-phonon lines in the spectra of color centers~\cite{pekar1950,huang1950}. Under these weak coupling conditions the ZRL becomes the most prominent feature in the spectrum. The corrections to the Rabi dynamics are expected to be small in this regime~\cite{dykman2025}. Therefore, achieving high two-qubit gate fidelities requires operation in the low pressing-field regime, necessitating careful tuning of the voltages applied to the underlying electrodes and adding a new parameter to consider in the optimization process.

A complete treatment of ripplon-induced dephasing requires a more rigorous analysis based on the Lindblad master equation, providing insight into the underlying physics across different parameter regimes~\cite{roy2011}. The ability to tune the coupling strength between electron motional states and the ripplonic bath over an energy scale comparable to the Coulomb interaction between two electrons provides a promising platform for investigating the interplay between dephasing and collective effects in coupled-electron systems~\cite{hall2025,quistrebert2025critical}.

\section{Conclusions}
\label{sec:conclusions}
In summary, we have shown that the entangling configurations from our previous work~\cite{beysengulov2024} can be utilized to achieve high-fidelity two-qubit gates, with the non-Clifford \sqiswap gate and the CZ gate as concrete examples. We have also shown that a direct minimization of the ZZ-coupling due to Coulomb interaction improves the fidelity for the \sqiswap gate compared to optimization for opposite-sign anharmonicities. On the other hand, it reduces the speed of the CZ gate due to the dynamics among the higher excited states. The exact cause of the reduced speed remains to be determined.

Our investigation of the stability of the gate fidelity reveals that even changes in ramp time smaller than a tenth of a nanosecond can negatively impact gate fidelity. For the \sqiswap gate, this impact can be reduced by minimizing the average energy difference between the first and second propagated eigenstates, $\ket{\Psi_1(t)}$ and $\ket{\Psi_2(t)}$, through the ramp-up stage of the gate. Hence, it is desirable to optimize the electrostatic double-well configurations such that these two states switch order between the idle and entangling stages of the gate. In the case of the CZ gate, we manage to achieve a stable gate performance with the \vzeta voltage function thanks to a strong suppression of swap-type interactions.

Although the scope of this study was limited to the motional states of the electrons, we emphasize the importance of our results also for electrons-on-helium qubits utilizing spin degrees of freedom. Given that current spin-state readout methods rely on spin-to-charge conversion techniques~\cite{schusterProposalManipulatingDetecting2010,zhangSpinorbitCouplingsDistant2012}, the thorough understanding of charge dynamics and electrostatic control presented here is essential not only for charge-based architectures but also as a foundation for spin-based approaches.

The theoretical insights presented in this work are particularly timely given recent experimental milestones. Strong coupling between the motional states of a single electron on helium in a quantum dot and microwave photons in a cavity has recently been demonstrated experimentally~\cite{koolstra2025strong}. It has also been shown that such quantum dots can be reliably loaded with two electrons, making the study of two-electron dynamics experimentally accessible. We anticipate that these developments, guided by the methodology presented here to minimize gate errors in a closed system, will enable future efforts to effectively isolate control-induced effects from environmental noise. Achieving this isolation is a prerequisite for understanding and mitigating decoherence, which is the next crucial step towards the first experimental realization of high-fidelity two-qubit gates with electrons on helium.

\begin{acknowledgments}
We would like to acknowledge Johannes Pollanen for helpful discussions.
OL has received funding from the European Union's Horizon 2020 research and
innovation program under the Marie Skłodowska-Curie grant agreement No. 945371. 
ZJS, JDW, and AKW acknowledge support from the U.S. Department of Energy, Office of Science, Basic Energy Sciences under Award Grant No. DE-SC0026211.
\end{acknowledgments}

\section*{Data Availability}
The data that support the findings of this article are openly available~\cite{leinonen_2025}.

\appendix

\section{Optimization of the \texorpdfstring{\vzeta}{Vζ} voltages}
\label{app:zz-opt}
The optimization of voltages for the \vzeta voltage function was primarily concentrated on directly minimizing the ZZ-interaction, $\zeta = E_4 - E_2 - E_1 + E_0$, for all configurations. In addition, the following loss terms were used when optimizing for specific configurations:

\begin{description}
    \item[Configuration I] One term to punish frequencies outside the 5-\SI{15}{\giga\hertz} operating range of the microwave cavities used for single qubit operations, and one term to punish qubit detuning smaller than \SI{3}{\giga\hertz}. These terms were weighted by a factor $10^{-2}$ relative to the $\zeta$ term.
    
    \item[Configuration II] One term to minimize $(E_2-E_1)^2$ to encourage an avoided crossing between the first and second excited states, which was weighted by a factor $10^{-4}$ relative to the $\zeta$ term. Additionally, two terms to punish energy differences between the third, fourth and fifth excited states smaller than \SI{1.5}{\giga\hertz} were used in order to keep these energy levels separated, hence suppressing the interaction among them. These two terms were weighted by a factor $10^{-2}$ relative to the $\zeta$ term.
    
    \item[Configuration III] Three terms for the triple degeneracy point between $\ket{\Phi_3}, \ket{\Phi_4}$ and $\ket{\Phi_5}$. The first two of these minimize the energy gaps $\Delta E_{54}^2 \equiv (E_5 - E_4)^2$ and $\Delta E_{43}^2 \equiv (E_4 - E_3)^2$, to enforce the degeneracy point. The third term minimizes $(\Delta E_{54}-\Delta E_{43})^2$ to symmetrize the two avoided crossings in the degeneracy point. These three terms had the same weight as the $\zeta$ term. Furthermore, the loss function included one term to punish energy differences between the first and second excited states below \SI{1.5}{\giga\hertz} in order to keep these energy levels separated and thereby suppressing the swap-type interaction. This term had a weight of $10^4$ relative to the $\zeta$ term.
\end{description}

\begin{figure*}
    \centering
    \includegraphics{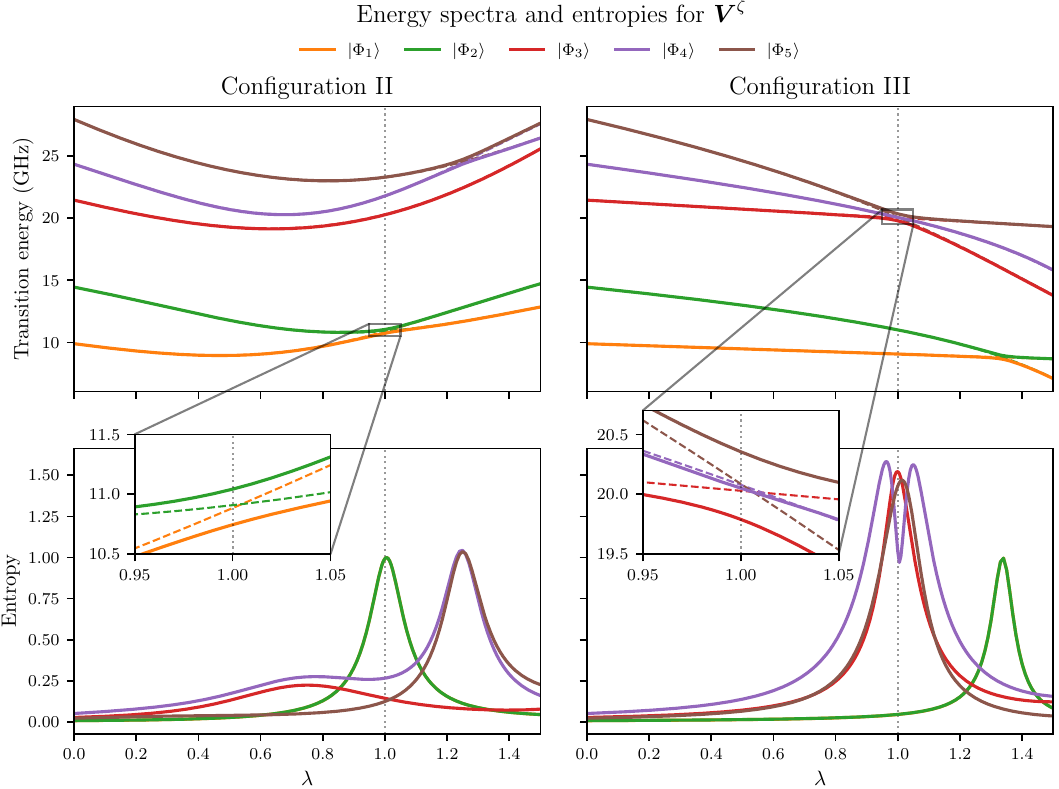}
    \caption{Transition energies from the ground state (top row) and von Neumann entropies (bottom row) of the five lowest excited energy eigenstates for $\vb*{V}^\zeta(\lambda) = (1-\lambda)\vb*{V}^\zeta_\mathrm{I} + \lambda\vb*{V}^\zeta_\mathrm{II/III}$, with $\vb*{V}^\zeta_\mathrm{II/III} = \vb*{V}^\zeta_\mathrm{II}$ (left) and $\vb*{V}^\zeta_\mathrm{II/III} = \vb*{V}^\zeta_\mathrm{III}$ (right). Dashed lines in the top plots represent the transition energies of the corresponding noninteracting systems~\cite{beysengulov2024}. The two relevant avoided crossing points are highlighted in the insets, where the energy gap between $\ket{\Phi_1}$ and $\ket{\Phi_2}$ for configuration II is \SI{300}{\mega\hertz} while the gap between $\ket{\Phi_3}$ and $\ket{\Phi_5}$ for configuration III is \SI{560}{\mega\hertz}. Note that the transition energies and entropies of the idle configuration I are included at $\lambda=0$ (in both the left and right plots), while the dotted vertical lines indicate $\lambda=1$, that is, the points where the system is kept during the hold time.}
    \label{fig:zz-spectra}
\end{figure*}

The weights of the different loss terms above were chosen heuristically by trial and error. We refer to our previous work \cite{beysengulov2024} for a more elaborate description of the optimization procedure. Optimization of these loss functions yielded the electrode voltage vectors $\vb*{V}^\zeta_\mathrm{I}, \vb*{V}^\zeta_\mathrm{II}$ and $\vb*{V}^\zeta_\mathrm{III}$ used in the \vzeta voltage function defined in Eq.~\eqref{eq:Vzeta}. Explicit values of the electrode voltages are reported in table~\ref{tab:voltages} below. The energy spectra and von Neumann entropies of the two resulting voltage functions (one for configuration II and one for configuration III) are presented in Fig.~\ref{fig:zz-spectra}. The energy gap between $\ket{\Phi_1}$ and $\ket{\Phi_2}$ for configuration II is \SI{297}{\mega\hertz}, while the gap between $\ket{\Phi_3}$ and $\ket{\Phi_5}$ for configuration III is \SI{565}{\mega\hertz}. A similar figure for the \vbeta voltage function can be found in Ref.~\cite{beysengulov2024}, where the energy gaps in the avoided crossing points have values of \SI{225}{\mega\hertz} and \SI{377}{\mega\hertz}, respectively.

\begin{table}[ht]
        \caption{Electrode voltage vectors $\vb*{V}^\zeta_\mathrm{I}, \vb*{V}^\zeta_\mathrm{II}$ and $\vb*{V}^\zeta_\mathrm{III}$ used in the \vzeta voltage function defined in Eq.~\eqref{eq:Vzeta}.}
        \begin{ruledtabular}
            \begin{tabular}{lccc}
                & $\vb*{V}^\zeta_\mathrm{I}$~[\unit{\milli\volt}] & $\vb*{V}^\zeta_\mathrm{II}$~[\unit{\milli\volt}]
                & $\vb*{V}^\zeta_\mathrm{III}$~[\unit{\milli\volt}] \\
                $V_1$ & 389.50 & 388.68 & 388.17
                \\
                $V_2$ & 200.70 & 206.69 & 194.01
                \\
                $V_3$ & 400.36 & 404.88 & 401.87
                \\
                $V_4$ & -290.61 & -288.37 & -289.10
                \\
                $V_5$ & 398.59 & 401.04 & 398.95
                \\
                $V_6$ & 200.15 & 192.98 & 198.82
                \\
                $V_7$ & 381.40 & 382.57 & 382.44
                \\
            \end{tabular}
        \end{ruledtabular}
        \label{tab:voltages}
    \end{table}

\section{Element-wise gate matrix analysis}
\label{app:gate_details}
\begin{figure*}[ht]
    \centering
    \includegraphics{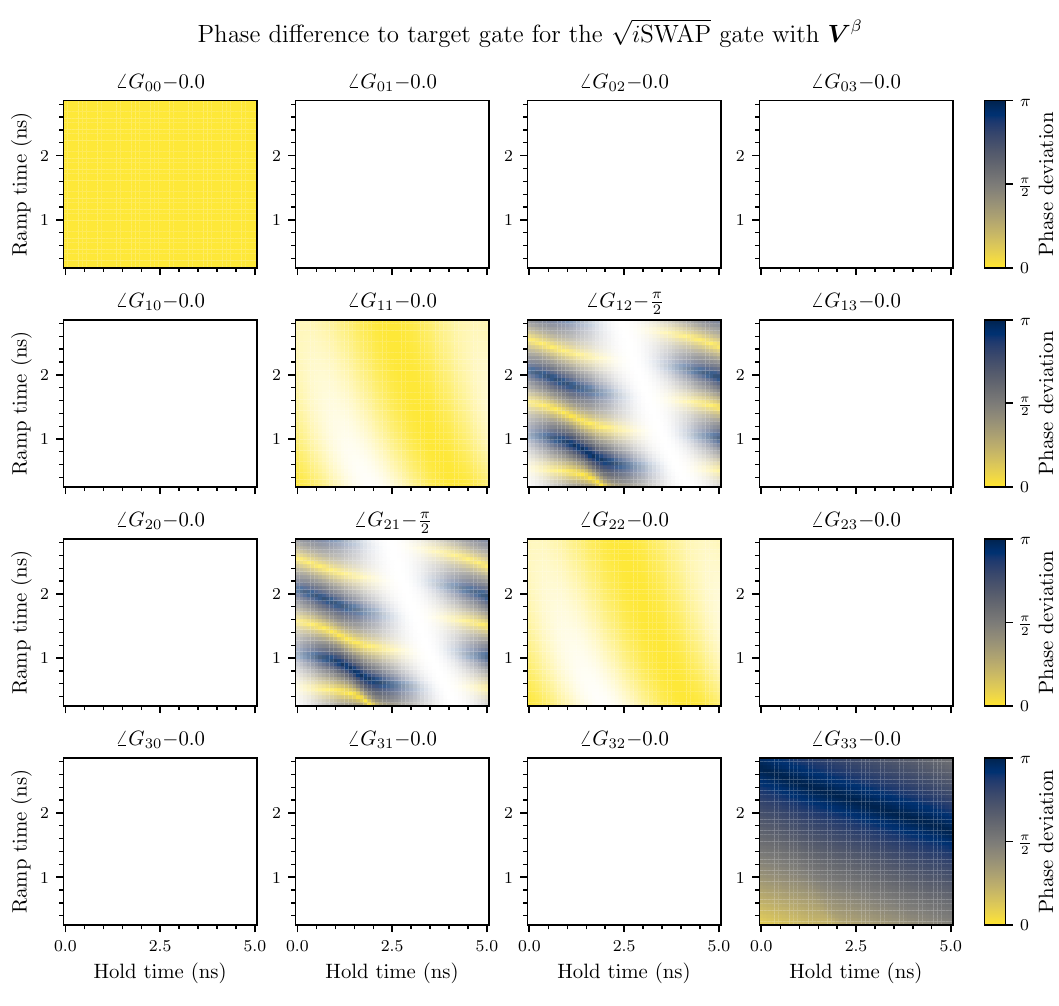}
    \caption{Element-wise analysis of the \vbeta \sqiswap gate. Each plot corresponds to one element $G_{ij}$ in the gate matrix $G(\theta\sbL,\theta\sbR)$, where the single qubit rotation angles $\theta\sbL$ and $\theta\sbR$ have been chosen such that $G_{ii}$ is real for $i\in\{0,1,2\}$. The color indicates the deviation from the ideal phase of the corresponding matrix element, as indicated in the respective plot titles, and the opacity signifies the amplitude. Note that the ideal amplitude, and thus opacity, of the four central plots is 0.5.}
    \label{fig:betaII}
\end{figure*}

\begin{figure*}[ht]
    \centering
    \includegraphics{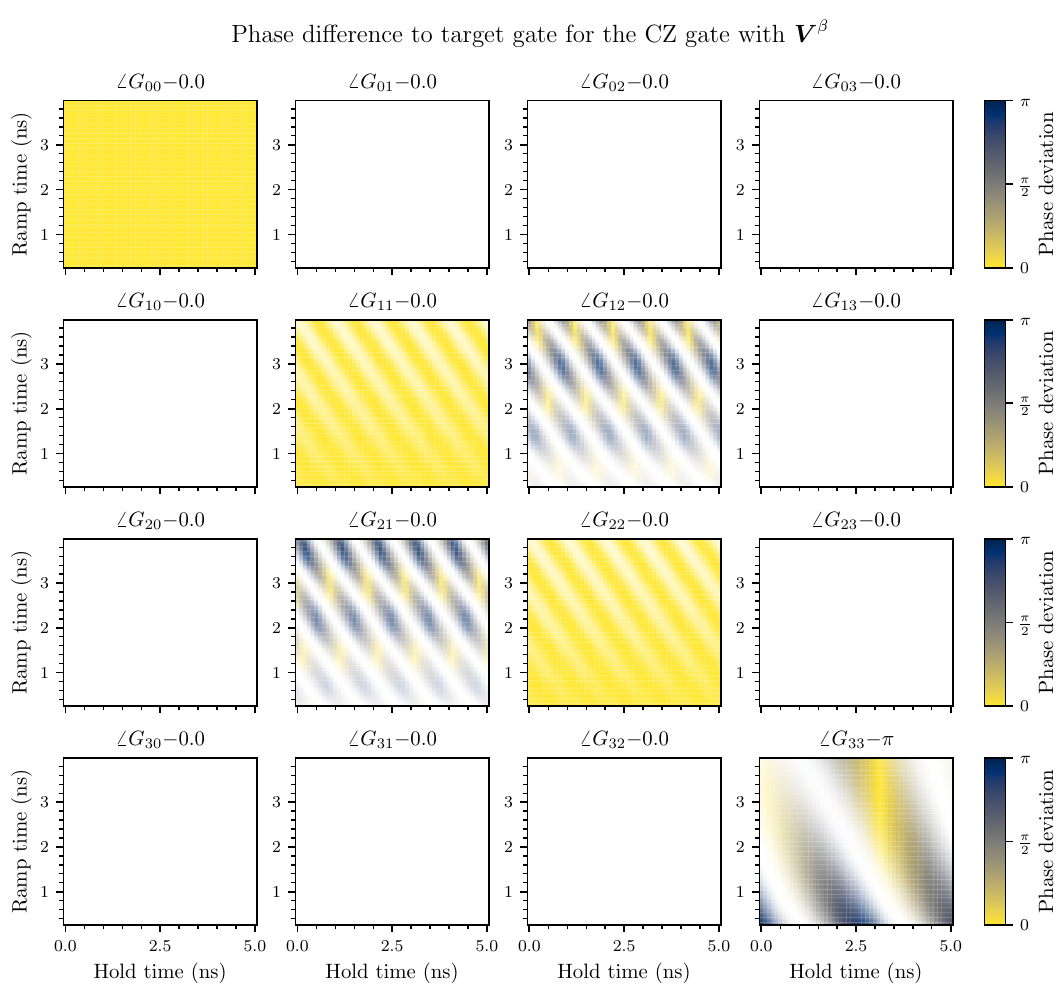}
    \caption{Element-wise analysis of the \vbeta CZ gate. Each plot corresponds to one element $G_{ij}$ in the gate matrix $G(\theta\sbL,\theta\sbR)$, where the single qubit rotation angles $\theta\sbL$ and $\theta\sbR$ have been chosen such that $G_{ii}$ is real for $i\in\{0,1,2\}$. The color indicates the deviation from the ideal phase of the corresponding matrix element, as indicated in the respective plot titles, and the opacity signifies the amplitude. The off-diagonal plots should ideally have opacity 0, i.e., be white.}
    \label{fig:betaIII}
\end{figure*}

\begin{figure*}[ht]
    \centering
    \includegraphics{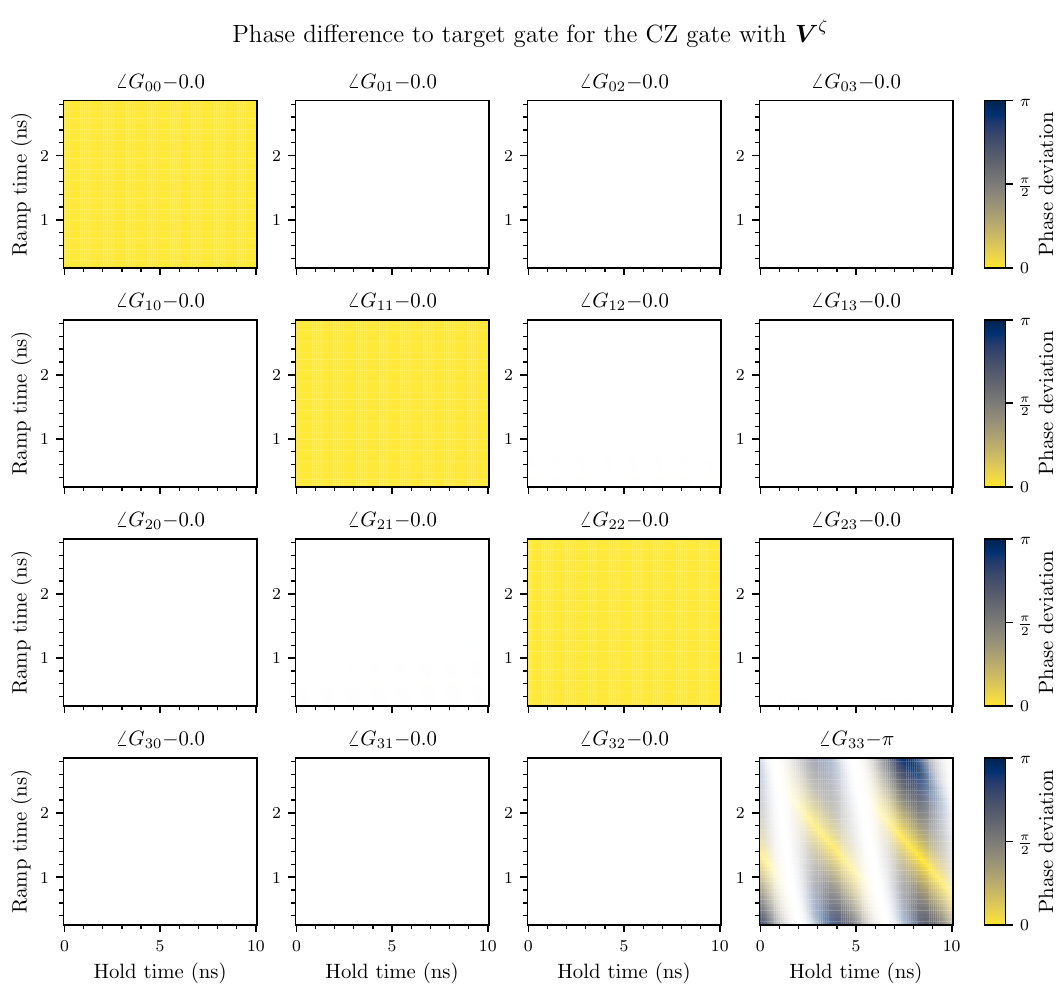}
    \caption{Element-wise analysis of the \vzeta CZ gate. Each plot corresponds to one element $G_{ij}$ in the gate matrix $G(\theta\sbL,\theta\sbR)$, where the single qubit rotation angles $\theta\sbL$ and $\theta\sbR$ have been chosen such that $G_{ii}$ is real for $i\in\{0,1,2\}$. The color indicates the deviation from the ideal phase of the corresponding matrix element, as indicated in the respective plot titles, and the opacity signifies the amplitude.}
    \label{fig:zetaIII}
\end{figure*}

In order to better understand the patterns that emerge in the gate fidelity, we will analyze the amplitude and phase of each matrix element separately in this section. We choose the single qubit rotation angles according to 
\begin{equation}
    \begin{aligned}
        \theta\sbL &= \angle U_{22} - \angle U_{00}\\
        \theta\sbR &= \angle U_{11} - \angle U_{00},
    \end{aligned}
    \label{eq:sq-ang}
\end{equation}
where $\angle U_{ii}$ denotes the complex phase of $U_{ii}$. This choice of angles collects the the phase of the diagonal elements of $G(\theta\sbL,\theta\sbR)$ on $G_{33}$, such that $G_{ii}$ is real for $i\in\{0,1,2\}$, thus allowing the general trends to emerge more clearly than they would with the optimized angles used in the main text. Throughout this section, we will adopt the notation $G_{ij}$ for matrix element $ij$ of the gate matrix $G(\theta\sbL, \theta\sbR)$ with $\theta\sbL, \theta\sbR$ given by Eq.~\eqref{eq:sq-ang}. The phase deviation and amplitude of these matrix elements are displayed in Fig.~\ref{fig:betaII} and Fig.~\ref{fig:zetaII} for the \sqiswap gate, and in Fig~\ref{fig:betaIII} and Fig.~\ref{fig:zetaIII} for the CZ gate, using \vbeta and \vzeta, respectively.

\begin{figure*}
    \centering
    \includegraphics{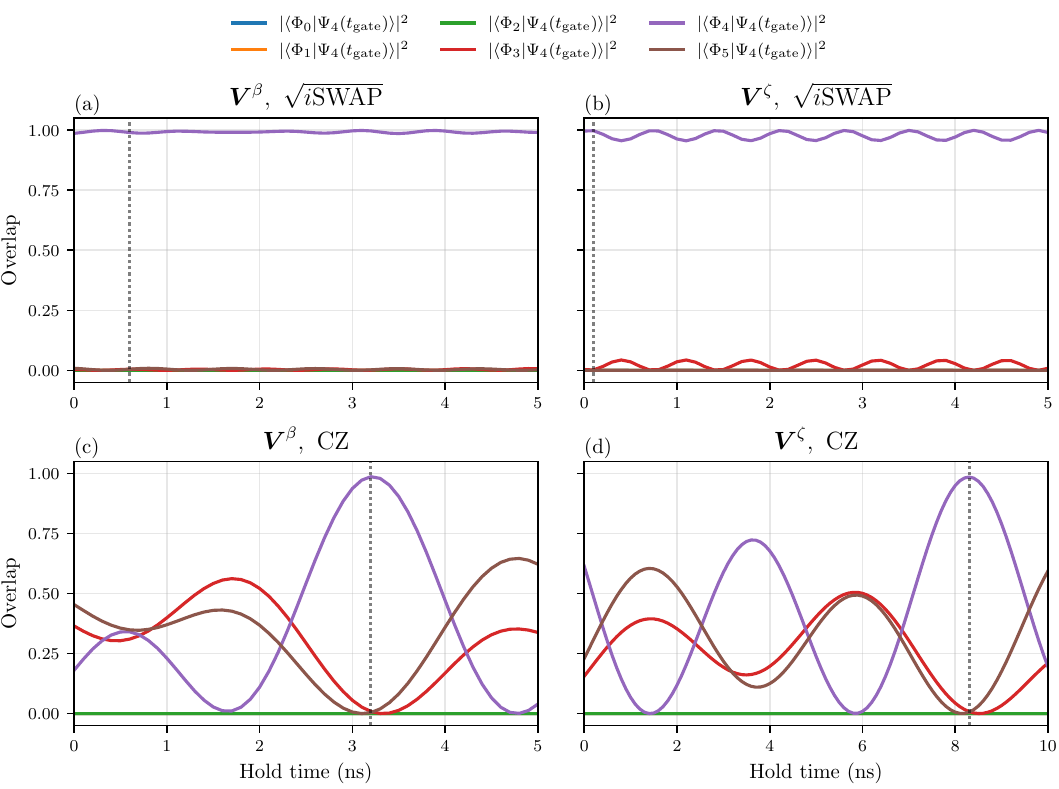}
    \caption{Overlap between the state $\ket{\Psi_4(\t{gate})}$, propagated by different two-qubit gate protocols, and the eigenstates $\ket{\Phi_n}$ before the gate protocol starts, as a function of hold time. All plots use the optimal ramp time for the respective gates and the optimal hold times are indicated by dotted vertical lines. The red and brown lines correspond to leakage from $\ket{11}$ to $\ket{02}$ and $\ket{20}$, respectively. For an ideal, leakage free, gate the overlap $\abs{\braket{\Phi_4}{\Psi_4(\t{gate})}}^2$ (purple line) should be 1.}
    \label{fig:33}
\end{figure*}

Figure~\ref{fig:betaII} shows the element-wise analysis of the \vbeta \sqiswap gate. By comparing with the fidelity in Fig.~\ref{fig:iswap-fid} (c), we see that the phase of the $G_{33}$ element is limiting high gate fidelity to small values of both ramp and hold time. However, the amplitude of this matrix element shows no dependence on the gate parameters, indicating that interactions among higher excited states are suppressed, which is also illustrated in Fig.~\ref{fig:33}~(a). Further structure in the fidelity stem from the phase and amplitude of the $G_{12}$ and $G_{21}$ elements.

The same data for the \vzeta \sqiswap gate is illustrated in Fig.~\ref{fig:zetaII}. We observe that the same matrix elements are crucial with this voltage function. The $G_{33}$ element is almost independent of the ramp time, likely because of the low $ZZ$-coupling in configuration~I, which is two orders of magnitude lower than in the \vbeta case. The small fluctuations in opacity are caused by interactions among the higher excited states, as shown in Fig.~\ref{fig:33}~(b). The phases of the $G_{12}$ and $G_{21}$ elements prove to be  crucial for the gate fidelity. The phase of both these elements oscillate with respect to the ramp time for both voltage functions, with a frequency of \SI{1.04}{\giga\hertz} for \vbeta and \SI{2.46}{\giga\hertz} for \vzeta. The frequency of these oscillations is closely related to the mean difference in energy between the $\ket{\Psi_1(t)}$ and $\ket{\Psi_2(t)}$ states during the ramp up from configuration~I to configuration~II, which is \SI{0.94}{\giga\hertz} for \vbeta and \SI{2.51}{\giga\hertz} for \vzeta. Since the same energy difference during the hold and ramp down stages of the gate operation is approximately \SI{0.1}{\giga\hertz}, and the phase is given by the integral of the energy over time, the phase difference of these matrix elements is dominated by the energy difference during the ramp up stage.

For the CZ gates, illustrated in Fig.~\ref{fig:betaIII} for \vbeta and Fig.~\ref{fig:zetaIII} for \vzeta, the fidelity is evidently mainly determined by the $G_{33}$ element. As discussed in the main text, the oscillations in amplitude of this matrix element stem from population leakage from $\ket{11}$ to the non-qubit states $\ket{02}$ and $\ket{20}$. Figure~\ref{fig:33} (c) and (d) show these oscillations as a function of hold time for \vbeta and \vzeta, respectively.  For the full fidelity of \vbeta, some additional complexity is added by the central elements, i.e. $G_{ij}$ with $i,j\in\{1,2\}$. This occurs because of an unwanted interaction between the $\ket{\Psi_1(t)}$ and $\ket{\Psi_2(t)}$ states, which stems from the passage through configuration~II when ramping the \vbeta function to configuration~III. This unwanted interaction is essentially eliminated with the \vzeta function since the parameterizations for the different gates are completely separated. 

\section{Full time evolution}
\begin{figure*}[ht]
    \centering
    \includegraphics{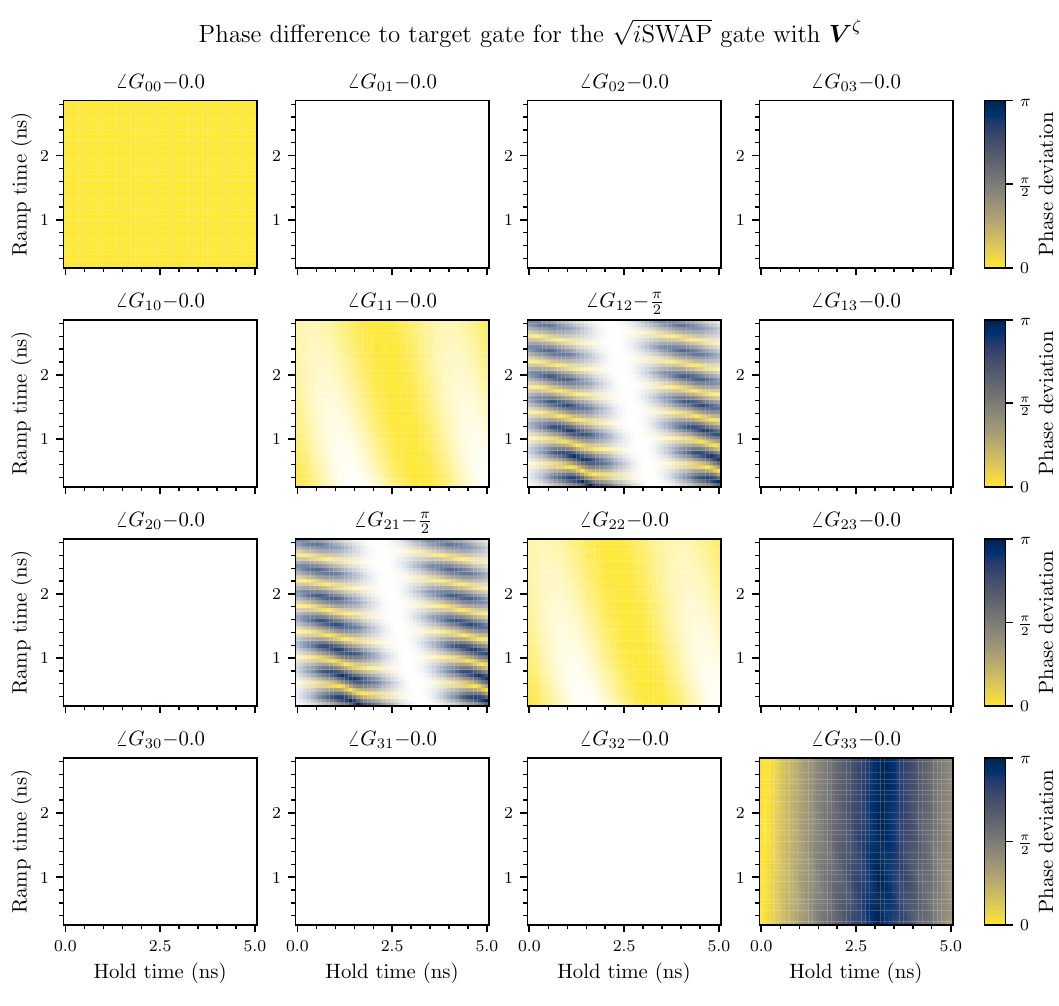}
    \caption{Element-wise analysis of the \vzeta \sqiswap gate. Each plot corresponds to one element $G_{ij}$ in the gate matrix $G(\theta\sbL,\theta\sbR)$, where the single qubit rotation angles $\theta\sbL$ and $\theta\sbR$ have been chosen such that $G_{ii}$ is real for $i\in\{0,1,2\}$. The color indicates the deviation from the ideal phase of the corresponding matrix element, as indicated in the respective plot titles, and the opacity signifies the amplitude. Note that the ideal amplitude, and thus opacity, of the four central plots is 0.5.}
    \label{fig:zetaII}
\end{figure*}

\begin{figure*}[ht]
    \centering
    \includegraphics{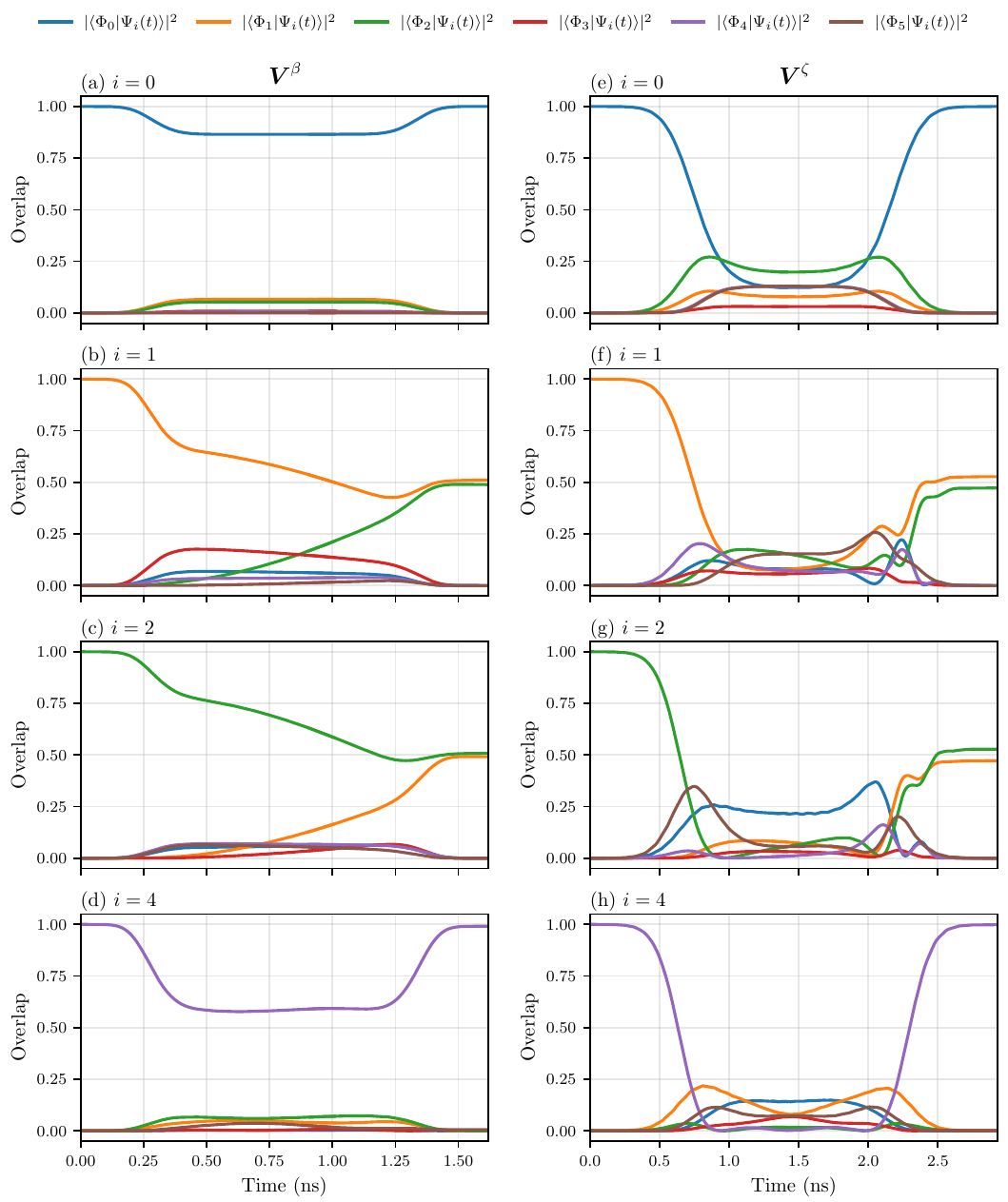}
    \caption{Time evolution of the overlaps between state $\ket{\Psi_i(t)}$ and the 6 lowest energy eigenstates at time $t=0$ during the optimal operation of the \sqiswap gate. The left column shows the dynamics using \vbeta (a--d) and the right column (e--h) shows \vzeta. Different subplots show different states $\ket{\Psi_i(t)}$, $i\in\{0,1,2,4\}$ as indicated by the subplot titles.}
    \label{fig:time-evolution-swap}
\end{figure*}

As stated in section~\ref{sec:method} of the main text, the computational basis states $\ket{00}, \ket{01}, \ket{10}, \ket{02}, \ket{11}$ and $\ket{20}$ are defined by the six lowest eigenstates $\ket{\Phi_n}$ at $t=0$, that is when the system is in configuration~I \cite{beysengulov2024}. Among them, the four relevant qubit states $\ket{00}, \ket{01}, \ket{10}$ and $\ket{11}$, i.e. the eigenstates $\ket{\Phi_0}, \ket{\Phi_1}, \ket{\Phi_2}$ and $\ket{\Phi_4}$, were propagated in time. In this section, we present the overlap $\lvert\braket{\Psi_i(t)}{\Phi_n}\rvert^2$ between the four propagated qubit states $\ket{\Psi_i(t)}, i\in\{0,1,2,4\}$, where $\ket{\Psi_n(0)} = \ket{\Phi_n}$, and the six computational basis states $\ket{\Phi_n}$, with $n\in\{0,\dots,5\}$, through the whole time evolution from $t=0$ to $t=\t{gate}$. In Fig.~\ref{fig:time-evolution-swap}, the dynamics for the \sqiswap gates are illustrated. We observe that all states become superpositions of the computational basis states when the system is in the interacting stage of the gate. This is not necessarily because of a population transfer, but rather because the configuration~I eigenfunctions do not constitute an eigenbasis for configuration~II. A completely adiabatic transition thus requires the eigenstates to become superpositions in the configuration~I eigenbasis. Consequently, the $\ket{\Psi_0(t)}$ states might at first glance appear to interact with other states, but the completely static composition during the hold stage of the gate reveals that this is not the case. The same is true for $\ket{\Psi_4(t)}$ for \vbeta, whereas small unwanted oscillations are observed for the same state in the case of \vzeta. These oscillations are also visible in the $G_{33}$ plot of Fig.~\ref{fig:zetaII}. They are however compensated for by an appropriate choice of hold and ramp time.

\begin{figure*}[ht]
    \centering
    \includegraphics{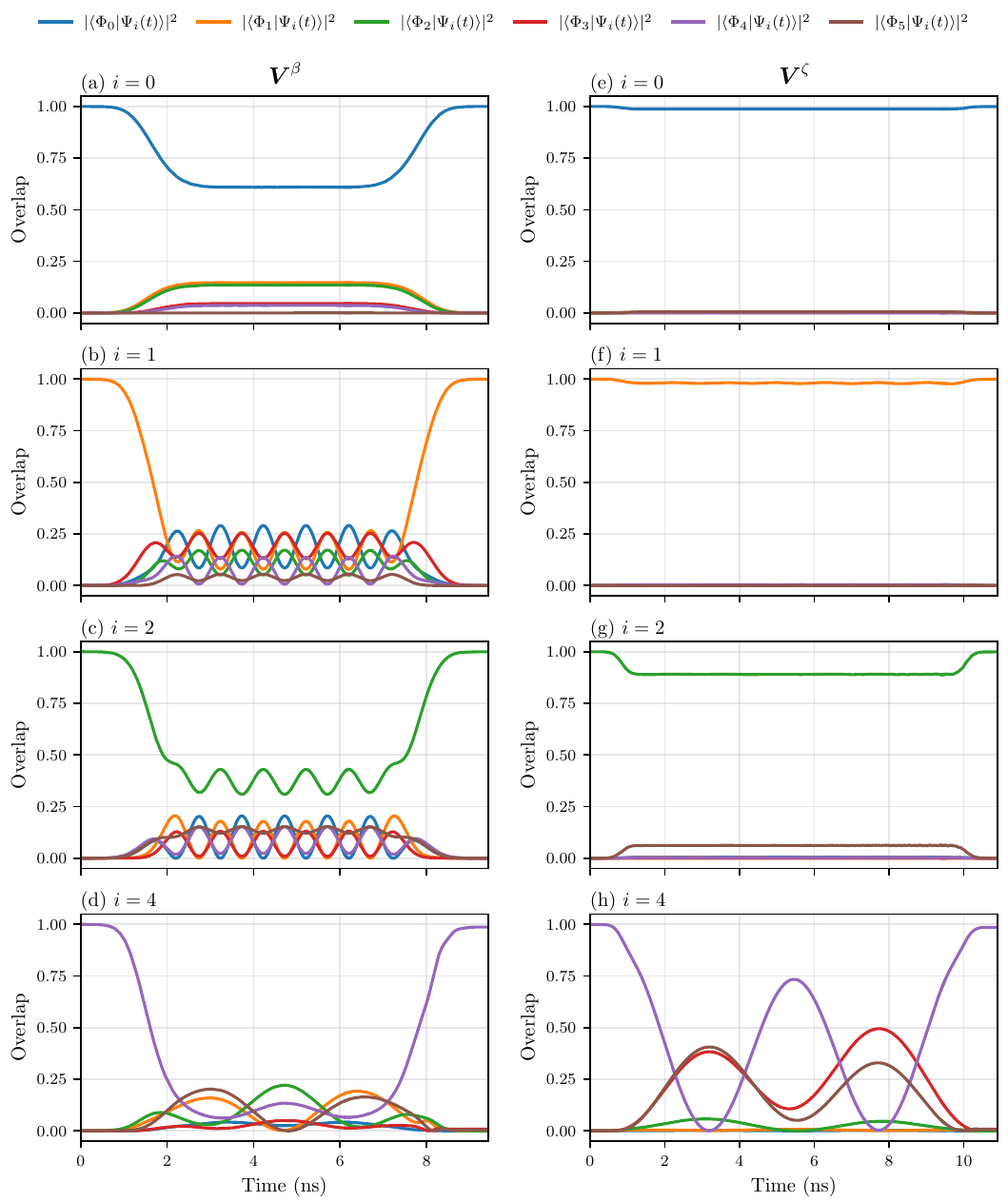}
    \caption{Time evolution of the overlaps between state $\ket{\Psi_i(t)}$ and the 6 lowest energy eigenstates at time $t=0$ during the optimal operation of the CZ gate. The left column shows the dynamics using \vbeta (a--d) and the right column (e--h) shows \vzeta. Different subplots show different states $\ket{\Psi_i(t)}$, $i\in\{0,1,2,4\}$ as indicated by the subplot titles.}
    \label{fig:time-evolution-cz}
\end{figure*}

In the case of the CZ gates, illustrated in Fig.~\ref{fig:time-evolution-cz}, the static behavior discussed above is desired for $0\leq i \leq 2$. This ideal behavior is observed for \vzeta as seen in plot (e)-(g), whereas \vbeta is subject to unwanted oscillations of the $\ket{\Psi_1(t)}$ and $\ket{\Psi_2(t)}$ states, illustrated in plot (b) and (c), respectively. As discussed in the main text and in appendix~\ref{app:gate_details}, this interaction is responsible for the sensitivity to ramp time observed for the \vzeta CZ gate.

\onecolumngrid

\clearpage
\twocolumngrid

\section{Screening of the Coulomb interaction}
\label{app:screening}

In order to estimate the impact of screening by the electrodes beneath the helium layer on gate time, we approximate the seven electrodes as an infinite metallic surface. Note that this is a significant simplification, as the empty spaces between the electrodes in the actual device are as wide as the electrodes themselves. This allows us to calculate the screened Coulomb interaction via the method of image charges~\cite{kawakami2023blueprint}. Since the Coulomb interaction depends solely on the relative distance between the electrons, we express the interaction as a function of the separation $r = |x_1 - x_2|$, denoted as $u(r)$. The resulting interaction, including both the direct repulsion between the electrons and the attraction between each electron and the image charge of the other, can be expressed as
\begin{equation}
    u(r) = \frac{\kappa}{r} - \frac{\kappa}{\sqrt{r^2 + (2h)^2}},
\end{equation}
where $h$ is the distance between the electrons and the conducting surface, and $\kappa$ is the Coulomb strength parameter defined in Eq.~\eqref{eq:hamiltonian}.

To determine the impact of screening on the gate speed, we perform a Taylor expansion of the interaction around the equilibrium electron separation $d$. Writing the instantaneous separation as $r = d + (\Delta x_2 - \Delta x_1)$, where $\Delta x_i$ is a small displacement of electron $i$, the expansion to second order yields
\begin{equation}
\mathtoolsset{multlined-width=0.85\displaywidth}
\begin{multlined}
    u(r) \approx u(d) + u'(d)(\Delta x_2 - \Delta x_1) \\ + \frac{1}{2}u''(d)(\Delta x_2 - \Delta x_1)^2.
\end{multlined}
\end{equation}
We note that the static term, $u(d)$, and the force terms proportional to $u'(d)$, only lead to shifts in the total energy and equilibrium positions, as they consist solely of single-particle terms where the electron displacements appear independently. While significant, these static shifts primarily affect the qubit frequencies and the resonance conditions for the gates. In our protocol, these effects can be compensated for by modifying the external potential through re-optimization of the voltage functions, $\vb*{V}(\lambda)$, to restore the necessary avoided crossings between energy levels.

Expanding the quadratic term reveals the operator $-u''(d)\Delta x_1 \Delta x_2$. This cross term is the only term in the expansion that couples the two electrons. Consequently, it is this term that drives the gate dynamics and determines the gate speed. We therefore focus on its strength, $u''(d)$. To quantify it, we define the screening factor $\eta$ as the ratio of the screened curvature to the unscreened curvature. The unscreened direct interaction $u_{\mathrm{dir}}(r) = \kappa/r$ has a curvature $u''_{\mathrm{dir}}(d) = 2\kappa/d^3$. For the image interaction term $u_{\mathrm{img}}(r) = -\kappa(r^2 + 4h^2)^{-1/2}$, the second derivative at $r=d$ is
\begin{equation}
    u''_{\mathrm{img}}(d) = -2\kappa \frac{d^2-2h^2}{(d^2 + 4h^2)^{5/2}}.
\end{equation}
The screening factor is then given by
\begin{equation}
    \eta = \frac{u''_{\mathrm{dir}}(d) + u''_{\mathrm{img}}(d)}{u''_{\mathrm{dir}}(d)} = 1 - \frac{d^3(d^2-2h^2)}{(d^2 + 4h^2)^{5/2}}.
\end{equation}
Defining the ratio $\xi = 2h/d$, this simplifies to
\begin{equation}
    \eta = 1 - \frac{1 - \xi^2/2}{(1 + \xi^2)^{5/2}}.
\end{equation}
For our device geometry, with $d=\SI{1.5}{\micro\meter}$ and $h=\SI{500}{\nano\meter}$, the expression above yields a screening factor of $\eta \approx 0.69$.

\section{Electron-Ripplon coupling Hamiltonian}
\label{sec:ehe-ripplon}

The electron coupling to the ripplon bath follows directly from the interaction Hamiltonian,
\begin{equation}
    \hat{\mathcal{H}}_i = \sum_{\mathbf{q}} V_{\mathbf{q}} e^{i\mathbf{q \cdot r}} \big ( b_{\mathbf{q}} + b_{-\mathbf{q}}^{\dagger} \big ),
    \label{eq:interaction}
\end{equation}
where $b_{\mathbf{q}}$ is the annihilation operator of a ripplon with wavevector $\mathbf{q}$, $\mathbf{r} = (x,y)$ is the electron coordinate, and $V_{\mathbf{q}}$ is the electron coupling parameter to the ripplonic field~\cite{schusterProposalManipulatingDetecting2010,platzmanQuantumComputingElectrons1999}. The coupling parameter is determined by a pressing field $E_{\perp}$, which is a controlled parameter in the experimental devices, and a polarization term $\mathcal{R}(q)$~\cite{monarkha2013two},
\begin{equation}
    V_{\mathbf{q}} = \sqrt{\frac{\hbar q}{2 \rho \omega_{\mathbf{q}}S_{\mathrm{He}}}} \Big (e E_{\perp} + \mathcal{R}(q) \Big ).
\end{equation}
Here $\omega_{\mathbf{q}} = \sqrt{\sigma\sbt{He}/\rho\sbt{He} q^3}$ is the ripplon dispersion relation, $\sigma\sbt{He}$ is the surface tension of liquid helium, $\rho\sbt{He}$ is the liquid helium density, and $S\sbt{He}$ is the helium surface area.
The dominant contribution to the interaction Hamiltonian arises from ripplons with wavevectors $q_x$ near the inverse electron localization length $l_x^{-1} = \sqrt{\hbar/m\sbt{e} \omega\sbt{e}}$, which is approximately \SI{40}{\nano\meter} for electron motional frequencies of $\omega\sbt{e}/2 \pi = \SI{10}{\giga\hertz}$. Typical frequencies of ripplons at these wavevectors, $\omega_{\mathbf{q}}/2\pi \approx \SI{0.4}{\giga\hertz}$, are small compared to $\omega\sbt{e}$. Therefore, single-ripplon electron motional state decay processes are strongly suppressed. However, ripplons can strongly contribute to dephasing by inducing fluctuations in the energies of the motional states~\cite{dykmanSpinDynamicsQuantum2023,dykman2025}. In the adiabatic approximation the energy modulation strength is characterized by a dimensionless parameter,
\begin{equation}
    \mathcal{C} = \sum_{\mathbf{q}} |\alpha_{\mathbf{q}}|^2 (2\bar{n}_{\mathbf{q}}+1),
    \label{eq:coupling_strength}
\end{equation}
where $\alpha_{\mathbf{q}} \approx V_{\mathbf{q}} / \hbar \omega_{\mathbf{q}}$ for relevant parameters $q_x l_x \sim 1$, and $\bar{n}_{\mathbf{q}}$ is the thermal occupation number of a ripplon with the wave number $\mathbf{q}$. The absorption spectrum linewidth $\gamma$ in the strong coupling regime is then given by
\begin{equation}
    \gamma = \Big(\sum_{\mathbf{q}} |\alpha_{\mathbf{q}}|^2 (2\bar{n}_{\mathbf{q}}+1) \omega_{\mathbf{q}}^2 \Big)^{1/2}.
    \label{eq:linewidth}
\end{equation} 

\newpage
\clearpage

\bibliography{references.bib}

\end{document}